\documentclass[twocolumn,showpacs,aps,prl]{revtex4}
\usepackage[usenames]{color}
\usepackage{graphicx}
\usepackage{amsmath}
\usepackage{amssymb}
\usepackage{bm}

\begin{document}

\title{Confined active nematic flow in cylindrical capillaries}

\author{Miha Ravnik}
\author{Julia M. Yeomans}

\affiliation{
Rudolf Peierls Centre for Theoretical Physics, University of Oxford, 1 Keble Road, Oxford OX1 3NP, UK}

\date{\today }

\begin{abstract}
We use numerical modelling to study the flow patterns of an active nematic confined in a cylindrical capillary, considering both planar and homeotropic boundary conditions. We find that active flow emerges not only along the capillary axis but also within the plane of the capillary, where radial vortices are formed. If topological defects are imposed by the boundary conditions, they act as local pumps driving the flow. At higher activity we demonstate escape of the active defects and flow into the third dimension, indicating the importance of dimensionality in active materials. We argue that measuring the magnitude of the active flow as a function of the capillary radius allows determination of a value for the activity coefficient.
\end{abstract}

\maketitle

There is currently widespread interest in understanding active systems which are key to the physics of life processes~\cite{sanchez11,koch11,sokolov07,riedel05}. Active systems are materials that produce their own energy, or respond to a continuous input of energy, so that they naturally operate out of thermodynamic equilibrium~\cite{ramaswamy10}. Examples range from bacterial suspensions~\cite{zhang10}, solutions of bio-filaments driven by molecular motors~\cite{koenderink09}, active colloids~\cite{dreyfus05,leoni08}, and cellular pattern formation~\cite{nedelec97}, to flocking birds and fish~\cite{cavagna09,katz11} and vibrated granular matter~\cite{kruelle09}.  Despite the diversity of these systems there is recurring behaviour, such as pattern formation~\cite{dombrowski04,giomi11}, the prevalence of instabilities~\cite{ramaswamy02} and the appearance of topological defects~\cite{vicsek95,toner95}. An important question in formulating theories of active matter is to identify the extent to which these behaviours are generic and to then identify the dominant control parameters.

Approaches that have proved very useful in categorising passive soft matter systems are continuum theories which rely primarily on the symmetries of the problems~\cite{simha02,kruse04,hatwalne04,luca12}. Several continuum theories have been proposed for active systems but matching theories to experiment, and giving a microscopic interpretation of the parameters, is still in its infancy~\cite{koch11}. Noting that nematic ordering and topological defects have been observed in active systems~\cite{sokolov07,wensink12,dogic12}, one of the most interesting candidates for a continuum approach is closely related to the Beris-Edwards equations of liquid crystal hydrodynamics~\cite{simha02,hatwalne04}. The important addition, representing the activity, is a term in the stress proportional to the nematic tensor order parameter. Hence any gradient in the order parameter field initiates a flow.

In the bulk, the activity drives instabilities which destroy the nematic order~\cite{mendelson99,simha02,saintillan07,wolgemuth08}. However, confining surfaces can stabilise a nematic state and corresponding time-independent flow fields. This has been demonstrated in a flat cell geometry~\cite{voituriez05,marenduzzo07}, in cells with static and moving boundaries~\cite{fielding11}. Other works have considered defect driven hydrodynamics in active circular regions surrounded by a passive  background~\cite{elgeti11} and spontaneous circulation in cytoplasmic streming~\cite{woodhouse12}.

In this Letter we consider confined active nematic flows in a cylindrical capillary, as a common example of an experimetnally relevant geometry. This is a very rich system as topological defects can emerge naturally, as a consequence of the boundary conditions. Solving the continuum equations of motion in the flow aligning extensile regime we find that active flow emerges not only along the capillary axis but also within the plane of the capillary, where radial vortices are formed. Interplay between the primary and secondary velocities leads to a spiralling flow field with a well-defined pitch moving down the capillary. We show that topological defects, imposed by heterogeneous boundary conditions, act as sources driving the flow.  We also argue that measuring the magnitude of the active flow as a function of the capillary radius allows determination of a value for the activity coefficient. 

Active nematic flow is approached by using numerical modelling based on continuum equations for apolar active materials~\cite{hatwalne04,marenduzzo07}. The model characterises the orientational order of active fluids by a symmetric and traceless order parameter tensor $Q_{ij}$, with the largest eigenpair representing the magnitude and direction of the local orientational order. In parallel, the flow of the active fluid is described by a velocity flow field $u_i$ and density $\rho$. The dynamic equations for $Q_{ij}$ and $u_i$ constitute the two governing equations of motion for the active nematic:
\begin{eqnarray}
(\partial_t+u_k\partial_k)Q_{ij}-S_{ij}&=&\Gamma H_{ij} +\lambda Q_{ij},\label{eqqten}\\
\rho(\partial_t+u_k\partial_k)u_i&=&\partial_j \Pi_{ij}+\eta\partial_{j}[\partial_i u_j +\partial_j u_i \nonumber \\
&+&(1-3\partial_\rho P)\partial_k u_k \delta_{ij}],\label{eqflow}
\end{eqnarray}
where $\partial_t$ is a derivative in time $t$, $\partial_i$ is a derivative in the Cartesian spatial coordinate $x,y,z$, $\Gamma$ is a rotational diffusion coefficient, $\eta$ is the viscosity, and $\lambda$ is a first active parameter which can be considered just as a renormalisation of the free energy. Summation over repeated indices is assumed. The leading contribution to the active dynamics is
controlled by the parameter $\zeta$ in the active contribution to the non-Newtonian terms of the stress tensor 
\begin{eqnarray}
\Pi_{ij}^{passive}&=&-P\delta_{ij}+2\xi(Q_{ij}+\delta_{ij}/3)(Q_{kl}H_{lk})\nonumber\\
&-&\xi H_{ik}(Q_{kj}+\delta_{kj}/3) -\xi (Q_{ik}+\delta_{ik}/3)H_{kj}\nonumber\\ &-&\partial_i Q_{kl}\frac{\delta \mathcal{F}}{\delta\partial_j Q_{lk}}+Q_{ik}H_{kj}-H_{ik}Q_{kj},\\
\Pi_{ij}^{active}&=&-\zeta Q_{ij},\label{eqActStress}
\end{eqnarray}
where $\xi$ determines whether the active material is flow tumbling or flow aligning. The relaxation to equilibrium of $Q_{ij}$ is determined by the molecular field $H_{ij}$, which originates from the equilibrium free energy $\mathcal{F}$ of an elastic anisotropic fluid, $H_{ij}=-\frac{\delta\mathcal{F}}{\delta Q_{ij}}+(\delta_{ij}/3)\mathrm{Tr}\frac{\delta\mathcal{F}}{\delta Q_{kl}}$, where:
\begin{eqnarray}
\mathcal{F}&=&L(\partial_k Q_{ij})^2/2\\
&+&AQ_{ij}Q_{ji}/2+BQ_{ij}Q_{jk}Q_{ki}/3+C(Q_{ij}Q_{jk})^2/4.\nonumber
\end{eqnarray} 
$L$ is the elastic constant, and $A,B,C$ are material constants.
Finally, the generalized advection term $S_{ij}$ and effective compressibility term $P$ are defined as:
\begin{eqnarray}
&&S_{ij}=(\xi D_{ik}+\Omega_{ik})(Q_{kj}+\delta_{kj}/3)\\
&&+(Q_{ik}+\delta_{ik}/3)(\xi D_{kj}-\Omega_{kj})-2\xi(Q_{ij}+\delta_{ij}/3)(Q_{kl} \partial_{k}u_l),\nonumber\\
&&P=P_0\rho-L(\partial_k Q_{ij})^2/2,
\end{eqnarray}
where $D_{ij}=(\partial_j u_i+\partial_i u_j)/2$, $\Omega_{ij}=(\partial_j u_i-\partial_i u_j)/2$ and $P_0$ is a constant related to the hydrostatic pressure. For more  details about the (non-active) model refer to~\cite{beris}.

\begin{figure}[!ht]
\centering \includegraphics[width=8cm]{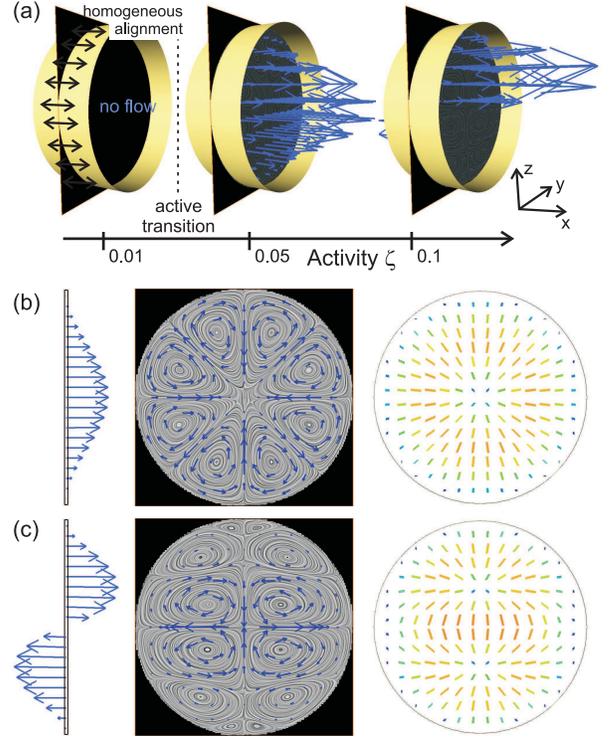}
\caption{Active flow in a capillary with homogeneous boundary conditions. (a) Flow regimes upon increasing activity: basic state with no flow, first excited state with unidirectional flow, and second excited state with bidirectional flow. (b,c) Primary flow $u_x$ across the capillary diameter; secondary flow in the $yz$ plane; and director profile of the active nematic (blue - out of plane, and red in plane) for (b) the unidirectional flow at $\zeta=0.05$, and (c) bidirectional flow at $\zeta=0.1$. } \label{Fig1}
\end{figure}

We solve the dynamic equations for the active flow $u_i$ and the orientation order $Q_{ij}$ by using a hybrid Lattice Boltzmann (LB) algorithm. The time-evolution of $Q_{ij}$ is obtained using an explicit finite difference scheme in time, whereas the equation for $u_i$  is solved by a D3Q15 lattice Boltzmann method~\cite{denniston04}. The simulations are performed in a circular simulation slice through a circular capillary tube with a cubic lattice; typically 4 x 200 x 200 mesh points. Along the capillary axis ($x$), we assume periodic boundary conditions for both $Q_{ij}$ and $u_i$, whereas at the capillary walls we take no-slip boundary conditions for $u_i$ (standard bounce-back) and fixed (in-plane or homeotropic) boundary conditions for $Q_{ij}$. All calculations are performed at small Reynolds numbers $Re$ (typically $Re\sim 10^{-5}$). The results are presented in units of the nematic correlation length $\xi$, the typical length scale over which the nematic order can vary, which allows for application of the results to different active systems. Typically, the correlation length $\sim$ a few times the length of the elementary constituents of the nematic and controls the characteristic size of the topological defects. For example, in bacterial suspensions~\cite{wolgemuth08}, $\xi$ $\sim 10~{\rm \mu m}$. Unless otherwise stated, the following values for the material parameters are used:
 $L =40~{\rm pN}$, flow aligining regime $\xi=1$, $\Gamma=7.29/{\rm Pa s}$, 
$A=-0.2L/\xi^2$, $B=-2.3L/\xi^2$, $C=1.9L/\xi^2$, mesh resolution $\Delta =1.5\xi$, time step  $\Delta t = 2.2 \xi^2/(L\gamma$), and capillary radius $R=150\xi$. These parameters and simulation regime would be appropriate for dense bacterial suspensions~\cite{wolgemuth08}, studied in microfluidic channels.


Figure~\ref{Fig1} shows the active flow profile in a capillary with surfaces imposing homogeneous alignment, parallel to the capillary axis $x$.  Upon increasing the activity (Fig.~\ref{Fig1}a), we observe a transition to spontaneous flow. Above the transition (first panel of Fig.~\ref{Fig1}b), the leading flow is along the axis of the capillary, ($x$ or $-x$), depending on the random noise in the inital condition for the director ($0.1\%$). Interestingly, at higher activities $\zeta$ (Fig.~\ref{Fig1}c), the flow evolves into a bidirectional pattern, with the velocity in opposite halves of the capillary in opposite directions.  The unidirectional flow can be interpreted as a  first excited active mode and the bidirectional flow as a second excited mode, selected by stronger active forcing. In is interesting to note that the stability of the two regimes partially overlap and they can be observed at the {\it same} activity $\zeta$. This indicates the existence of energy barriers between active flow states. Indeed, analogous flow patterns and metastability of flow regimes was also observed in simulations of two-dimensional active nematics~\cite{orlandini08}.  

\begin{figure}[ht]
\centering \includegraphics[width=8cm]{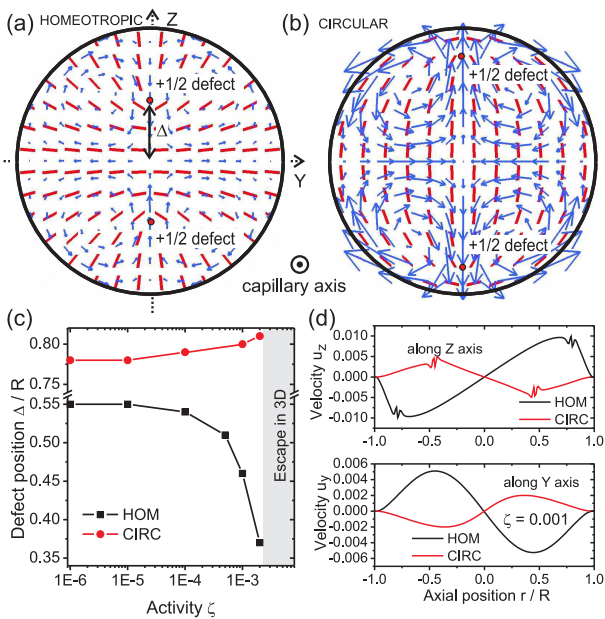}
\caption{Active $+1/2$ defects are stabilised by cylindrical confinement in a capillary. Active flow (in blue) and director (in red) for (a) homeotropic (perpendicular) and (b) circular director alignment at the capillary walls. The director and flow lie fully in the $(y,z)$ plane. (c) Position of the defects $\Delta$ as a function of the activity $\zeta$ for homeotropic and circular   surface. (d) Velocities $u_y$ and $u_z$ for homeotropic and circular surface at $\zeta=0.001$ along the $y$ and $z$ axes, respectively.} \label{Fig2}
\end{figure}

However the full three dimensional simulations also allow us to identify an inplane, secondary flow field within the $yz$ capillary cross-section (Fig.~\ref{Fig1}b and c, second panels). The secondary flow consists of a circularly symmetric pattern of distinct vortices, 8 or 12 for the unidirectional and bidirectional flows, respectively. The vortices are pair-wise counter-rotating, thus exhibiting no net angular momentum.  The magnitude of the secondary flow is $\sim 1\%$ of that of the primary flow.
The complex structure of the flow is reflected in the active nematic director, shown in Fig.~\ref{Fig1}b and c, third panels. The director exhibits a  variable splay deformation in the radial direction of the capillary and breaks the cylindrical symmetry in the bidirectional flow. Interestingly, the observed primary and secondary active flows are separately analogous to the flow phases predicted in active polar films~\cite{voituriez06}. 

Topological defects in the orientational profiles of active materials could be used as sources of the microscopic activity, and a cylindrical capillary can naturally support such active defect profiles. Two active defect configurations are shown in capillaries with surfaces imposing homeotropic (Fig.~\ref{Fig2}a) or circular (Fig.~\ref{Fig2}b) director alignment. In both configurations, the two defect regions appear as a result of the orientational order being frustrated by the confining surfaces and are actually defect lines along the capillary with $+1/2$ winding number. Surrounding the defects, a strictly in plane ($yz$) active flow develops  with zero flow component along the capillary axis. The magnitude of the flow is largest in the region surrounding the defects, which indicates that the defects act as effective local 'pumps' for the flow. It is the strong elastic distortion and variation in the degree of order in the defect cores, which produces locally high stress and consequently strong active flow. The small structures in the velocity field at the position of the defects seen in Fig.~\ref{Fig2}d occur because local changes in the magnitude of the order parameter also cause elastic and active stresses. Flow along the direction of the defects ($z$) is larger by a factor of $\sim 2$ than the flow in the perpendicular direction ($y$). Note that the orientation of the defects  clearly determines also the \textit{direction} of the local active flow (compare Fig.~\ref{Fig2}a and Fig.~\ref{Fig2}b). And depending on this direction, the defects are advected by the flow either towards the centre or the walls of the capillary (Fig.~\ref{Fig2}c). If defects approach the surface, they produce larger distortions in the nematic and thus generate stronger flows. Quantitative flow profiles along the $y$ and $z$ capillary axes are shown for the homeotropic and circular boundary anchoring at $\zeta=0.001$ in Fig.~\ref{Fig2}d.  Finally,  although the calculations are performed in a full three-dimensional capillary, the specific configuration with defects in Fig.~\ref{Fig2} can be interpreted as quasi-two-dimensional active nematic  in a circular confinement, since both flow and director have a strictly zero component along the capillary axis. 

\begin{figure}[!ht]
\centering \includegraphics[width=7.5cm]{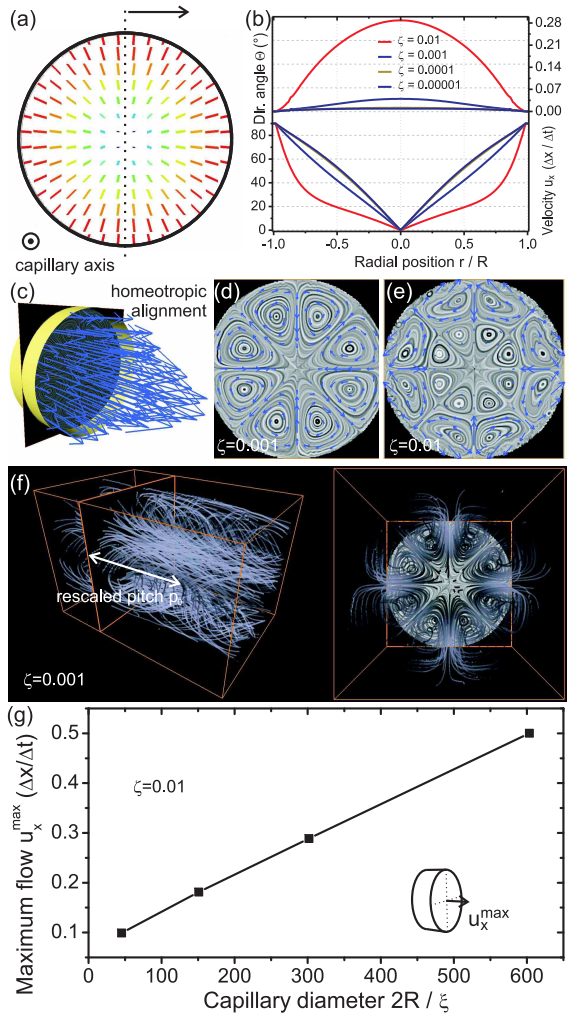}
\caption{Escape of the active flow in third dimension. (a) Escaped active nematic director (red is in-plane, blue out-of-the-plane). (b) Velocity and director angle profiles across the capillary. (c) Primary and (d,e) secondary active flow profiles escaped along the capillary axis. (f)  Stream-lines showing full active flow reveal an array of helical vortices. For visualisation, the pitch is rescaled by a factor of 50 and data is drawn for 100 simulation boxes, which were periodically multiplied along the capillary axis. (g) Maximum flow in the centre of the capillary at $\zeta=0.01$ as a function of the capillary radius.} \label{Fig3}
\end{figure}

At large activities, the active flow and the active defects escape in the third dimension (along the capillary axis), as indicated in Fig.~\ref{Fig2}c and demonstrated in Fig.~\ref{Fig3}. The director flips out of the $yz$ plane, and thus accommodates the distortion imposed by the surface by escaping in the third dimension rather than with the two $+1/2$ defects. The active flow simultaneously develops a  primary $x$-component along the capillary axis (Fig.~\ref{Fig3}c). For $\zeta=0.001$, the ratio between the maximum secondary flow $u_{yz}^{max}$ and maximum  $x$-component of the flow  $u_{x}^{max}$ is $u_{yz}^{max}/u_{x}^{max}=0.013$. The nematic order is distorted and therefore there is no transition back to a zero flow state or an in plane director configuration upon reducing the activity back to zero. Thus the escaped flow regime can also be stable at the same lower activities ($\zeta \to 0$) where defect states were observed (Fig.~\ref{Fig2}), which is an indication of (meta)stable director fields -and consequently flow- states with energy barriers. 

The escaped active flow has, as for the homogenous capillary, a complex in-plane secondary flow. A lattice of vortices forms, with the number of vortices depending on the activity: 8 vortices for $\zeta=0.001$ (Fig.~\ref{Fig3}d) and 16 vortices for $\zeta=0.01$ (Fig.~\ref{Fig3}e). Figure~\ref{Fig3}f depicts the primary flow along $x$ direction and the  secondary flow in the $yz$ plane by showing flow streamlines. This reveals helical flow motifs, i.e. a lattice of vortices with a distinct pitch. The pitch of the active vortices is $p_0\approx 11000\xi$ for $\zeta=0.001$. The helical flow is reminiscent of cytoplasmic streaming in plant cells~\cite{goldstein08}. 

We next present results to determine the scaling of the active flow with the physical size of the capillary. Figure~\ref{Fig3}g shows the maximum flow velocity $u_{x}^{max}$, in the centre of the capillary and along the capillary axis $x$, for various diameters of the capillary $2R$ at a given $\zeta=0.01$.
The dependence $u_{x}^{max}(R)$ is to good approximation linear,  indicating a predominantly Poiseuille-like flow regime for the escaped active flow. The maximum velocity of a general isotropic fluid in a circular channel scales as $u_{x}^{max}|_{iso}=fR^2/4\eta$, where $f$ is an effective body force driving the flow, e.g. a pressure drop per unit length of a capillary $f=\Delta p/L$. Indeed, the active contribution in the generalised Navier-Stokes equation~(\ref{eqflow}) and (\ref{eqActStress}) can be extracted as an effective force from the active stress tensor  $f_i=\partial_j \Pi_{ij}^{active}=-\zeta \partial_j Q_{ij}=-(\zeta/R) \hat\partial_j Q_{ij}\propto (\zeta/R) $, where assuming scalalability of $Q_{ij}$ the scale independent coordinates $\hat x_i=x_i/R$ have been introduced. By writing $u_{x}^{max}\propto (\zeta/R)R^2/4\eta\propto R$, the linear scaling of $u_{x}^{max}$ is obtained, in agreement with the simulation results in Fig.~\ref{Fig3}g. Importantly, the escaped active nematic orientational profile is well scalable in space and thus gives linear $u_{x}^{max}\propto R$, which is not necessarily true for other active nematic geometries or profiles. Note also that the slope of the $u_{x}^{max}(R)$ curve is proportional to  $\zeta/\eta$, and could be used to measure this ratio in active nematic materials.  Indeed using an alternative viscosity measurement the strength of the activity could also be determined, and the full relation between $u_{x}^{max}$ and $\zeta/\eta$ (not only the proportionality) could be found by measuring at different $\zeta/\eta$ ratios, e.g. by changing the temperature, yet retaining the escaped nematic profile.

As experiments on active material in microchannels are increasingly feasible, there are several possible applications for our results. Flow patterns similar to those seen here may emerge in dense suspensions of swimmers confined to microfluidic channels which have a preferred orientation relative to the surface~\cite{berke08}. Dense suspensions of active colloids provide another possible relevant experimental system~\cite{theurkauff12}. Moreover similar flow patterns are seen in systems of collectively moving microtubules~\cite{sumino12}.

In conclusion, we have demonstrated the richness of active nematic flow and orientational profiles in a cylindrical capillary tube. The physical behaviour depends strongly on the surface boundary conditions imposed on the director field: we considered homogenous planar, circular planar, and homeotropic (perpendicular) anchoring. The active flow emerged above a threshold in the activity for intrinsically  homogeneous orientational profiles, but at $\zeta = 0$ for intrinsically inhomogenous director fields. The primary active flow developed along the capillary, but with  intricate secondary flows, consisting of an array of velocity vortices. At high activities, bidirectional primary flow was found, indicating multiple activity modes which are observed to be (meta)stable relative one to another. Topological defects imposed by the boundary conditions in this extensile active nematic study were shown to act as effective local pumps for the flow, with their topological structure determining the pumping direction. We demonstrated the escape of the active flow and active defects into the third dimension, showcasing the importance of the dimensionality in active materials, and providing a possible route to measuring the activity of active suspensions.   

M. R. acknowledges support of the EU under the Marie Curie Programme ACTOIDS.

\end{document}